\documentclass[12pt,reprint,tightenlines,groupedaddress,superscriptaddress,floatfix,aps]{revtex4}
\usepackage{graphicx}
\usepackage{rotating}
\usepackage{array}
\usepackage{multirow}
\usepackage{epstopdf}
\usepackage{amsmath}
\usepackage{amsfonts}
\usepackage{setspace}
\usepackage{braket}
\usepackage{color}

\usepackage{hyperref}

\begin{document}
\title{Evaluation of Neel temperatures from fully self-consistent broken-symmetry GW and high-temperature expansion: application to cubic transition-metal oxides}
\author{Pavel Pokhilko}
\affiliation{Department  of  Chemistry,  University  of  Michigan,  Ann  Arbor,  Michigan  48109,  USA}
\author{Dominika Zgid}
\affiliation{Department  of  Chemistry,  University  of  Michigan,  Ann  Arbor,  Michigan  48109,  USA}
\affiliation{Department of Physics, University of Michigan, Ann Arbor, Michigan 48109, USA }

\renewcommand{\baselinestretch}{1.0}

\begin{abstract}
Using fully self-consistent thermal broken-symmetry GW we construct effective magnetic Heisenberg Hamiltonians for a series of transition metal oxides (NiO, CoO, FeO, MnO), capturing a rigorous but condensed description of the magnetic states. 
Then applying high-temperature expansion, we find the decomposition coefficients for spin susceptibility 
and specific heat. 
The radius of convergence of the found series determine the Neel temperature. 
The NiO, CoO, and FeO contain a small ferromagnetic interaction between the nearest neighbors (NN) and the dominant antiferromagnetic interaction between the next-nearest neighbors (NNN). 
For them the derived Neel temperatures are in a good agreement with experiment.  
The case of MnO is different because both NN and NNN couplings are antiferromagnetic and comparable in magnitude, 
for which the error in the estimated Neel temperature is larger, which is a signature of additional effects not captured by electronic structure calculations. 
\end{abstract}
\maketitle

Simulating and designing novel antiferomagnetic materials (AFM) materials has been one of the main interests of solid-state chemistry, condensed matter, materials science, as well as engineering.  
AFM are of technological importance for 
designing spin valves, room-temperature electrical switches, colossal magnetoresistive effects,  fast magnetic moment dynamics, and other antiferromagnetic spintronic applications. 

The Neel temperature is the most important characteristic parameter describing AFM. It determines their technological applicability, since above the Neel temperature AFM undergoes a phase transition to a disordered phase from a phase where spins are displaying antiferromagnetic ordering. 
The theoretical determination of Neel temperature for realistic systems is very challenging due to several reasons.  
First, electronic structure of AFM is multiconfigurational and requires simultaneous treatment of both strong and weak electron correlation. 
Second,  according to the Hohenberg--Mermin--Wagner theorem,  magnetic critical phenomena can happen only in 3D bulk. Consequently, numerical simulations should be performed for periodic problems with taking into account possible finite-size effects. This requirement limits techniques such as exact diagonalization that are traditionally used for model Hamiltonians. 
Moreover, even for paradigmatic model systems such as 3D Hubbard model,  a brute-force diagrammatic approach of evaluating finite-temperature susceptibilities gave unsatisfactory estimates of critical temperatures\cite{Iskakov:phase_tr:2022}. 
Not surprisingly, given the aforementioned challenges, there is a lack of rigorous quantum estimates of Neel temperatures for realistic AFM. 
For example, Neel temperatures of the transition-metal oxides studied in this work were estimated 
only with the techniques based on a classical Heisenberg model\cite{Majumdar:NiO:MnO:DFT:J:2011,Savrasov:LDA:DMFT:Neel:T:2006} with somewhat ambiguously parametrized density functionals and a Hubbard $U$.

While using only model systems or brute-force realistic diagrammatic approaches have severe shortcomings, an alternative approach that is ab-initio and is based on quantum mechanics is still viable. 
In such an approach, 
effective exchange couplings present in the quantum Heisenberg Hamiltonian can be extracted 
from DFT or wave-function calculations
only using a limited number of computed states. These states are routinely accessible in standard electronic structure calculations.
Subsequently, the constructed quantum Heisenberg Hamiltonian is extrapolated to encompass the states that cannot be efficiently captured by electronic structure calculations. Such a procedure is a very efficient way of computation of magnetic properties for molecular systems\cite{Mayhall:2014:HDVV,Mayhall:1SF:2015,Pokhilko:EffH:2020,Pokhilko:spinchain,Kotaru:Fe:SMM:2022}. 
In the past, Pokhilko applied this extrapolation within equation-of-motion spin-flip coupled-cluster theory\cite{sfpaper} not only for the states with low spin projections that cannot be captured, but also to the infinite system size, 
fully reproducing experimental magnetic susceptibility without any adjustable parameters\cite{Pokhilko:spinchain}.

Recently, we introduced a broken-symmetry GW approach~\cite{Pokhilko:local_correlators:2021,Pokhilko:BS-GW:solids:2022} based on the finite-temperature self-consistent GW code~\cite{Iskakov20,Yeh:GPU:GW:2022} and benchmarked it for molecular systems and solids. 
In this work, we further extend the idea of using broken-symmetry GW to construct effective Hamiltonians according to the scheme below:
\begin{enumerate}
\item Extraction of magnetic couplings $J$ from wave-function or broken-symmetry self-consistent periodic GW calculations and 
estimation of finite-size effects for $J$. 
\item Evaluation  of an extrapolated Heisenberg Hamiltonian for a large or even infinite system and 
reconstruction of the magnetic manifold of states of the target system.  
\item Application of a high-temperature expansion (HTE) for spin susceptibilities and specific heat. 
\item Determination of convergence radii of the expansions yielding the estimates of the Neel temperature. 
\end{enumerate}
We benchmark this approach on bulk NiO, CoO, FeO, and MnO against other theories and experiments. 

\textbf{Effective Hamiltonian.} 
Magnetic phenomena can be accurately described with effective Hamiltonians since magnetic states are usually separated well from other states in the spectrum.
In wave-function theories, such effective Hamiltonians can be constructed within the Bloch formalism as a result of an exact transformation from target to model spaces\cite{Calzado:02,Guihery:2009,Malrieu:2010,Marlieu:MagnetRev:2014,Mayhall:2014:HDVV,Mayhall:1SF:2015,Pokhilko:EffH:2020,Pokhilko:spinchain}.  
However, in Green's function or density-based methods, the wave-function amplitudes cannot be accessed directly and an alternative strategy has to be employed. This strategy relies on using broken-symmetry solutions possible to access in these methods. These Ising-like broken-symmetry solutions are found and their energies are used 
to extract magnetic couplings\cite{noodleman:BS:81,Yamaguchi:BS:formulation:1986}. 
In this paper to find these solutions, we use the fully self-consistent, finite-temperature GW approach starting from the unrestricted Hatree--Fock guess.  The details of the broken-symmetry GW strategy can be found in Ref.~\cite{Pokhilko:BS-GW:solids:2022}. 

For the metal oxides, we consider the following form of the the effective Hamiltonian established by previous broken-symmetry DFT calculations\cite{Martin:NiO:exchange:2002,Feng:MnO:CoO:b3lyp:2004,Kresse:MnO:PBE:2005,Majumdar:NiO:MnO:DFT:J:2011} 
\begin{gather}
H = -J_{1} \sum_{\braket{i,j}} \vec{S}_i \vec{S}_j - J_{2} \sum_{\braket{\braket{i,j}}}  \vec{S}_i \vec{S}_j,
\protect\label{eq:Ham_defs}
\end{gather}
where $\braket{i,j}$ are the unique nearest-neighbor (NN) pairs, 
$\braket{\braket{i,j}}$ are the unique next-nearest-neighbor (NNN) pairs, 
$J_1$ is the NN effective exchange coupling, 
$J_2$ is the NNN effective exchange coupling, 
$\vec{S}_i$ are the local (model) spins on metal center $i$.  
In all compounds studied in this work, the $J_2$ constant is antiferromagnetic. 
Therefore, all these oxides show spin frustration that makes the determination of 
critical point especially difficult. 
To obtain exchange couplings, we use energy differences between the solutions with maximum and zero (broken-symmetry) 
spin projections evaluated using different unit cells. 
The exchange couplings are extracted according to the expressions below
\begin{gather}
J_{1,u} = -\frac{E(HS1)-E(BS1)}{16}, \\
J_{2,u} = -\frac{E(HS2)-E(BS2)}{12} - J_{1,u},  
\protect\label{eq:J_extr}
\end{gather}
where $E(\cdots)$ denotes the energy of the high-spin (HS) and broken-symmetry (BS) solutions in cells 1 and 2 (see Ref. \cite{Pokhilko:BS-GW:solids:2022} for details). 

In this work, we neglect relativistic effects since according to the Kanamori's estimate\cite{Kanamori:FeO:CoO:I:1957} the impact of the spin--orbit splitting in FeO and CoO on the Neel temperature is negligible. This is because the states, which as a result of spin-orbit coupling are non-degenerate,  are fully thermally accessible.
Our non-relativistic calculations when compared with experimental data will provide a good test of this prediction. 
While in this work, the impact of the relativistic effects on the Neel temperature is not significant, in general, the spin--orbit couplings between the degenerate non-relativistic states in Fe(II) and Co(II) are expected to be strong~\cite{Kanamori:FeO:CoO:I:1957} and generalized in a more mathematically precise way by Pokhilko\cite{Pavel:SOCNTOs:2019}, resulting in spin--orbit splitting and single-ion anisotropy. This interplay of correlation and relativistic effects is of interest to both experimental and theory groups~\cite{Stock:CoO:2020}.   

\textbf{High-temperature expansion (HTE).} 
HTE was introduced by Opechowski in 1937\cite{Opechowski:1937} and was subsequently applied to various Heisenberg and Ising Hamiltonians\cite{Rushbrooke:Wood:HTE:1955,Wood:Rushbrooke:HTE:1957,Rushbrooke:Wood:Curie:1958,Domb:Sykes:HTE:Curie:1957,Young:triang:1993,Young:kagome:1994}. 
A prescription for a most general form of HTE for several coupling constants and arbitrary spin has only been  introduced recently\cite{Richter:HTE8:2011,Richter:HTE10:code:2014}. 
The benefit of HTE is its applicability to highly frustrated systems.  
The main advantage of HTE is its a relatively low computational cost of evaluation of its coefficients. This allows one to study real space cells large enough to reach thermodynamic limit (TDL), thus eliminating the dependence of the final answer on the finite-size effects.

The idea of the HTE is the following. 
Thermodynamic quantities, such as spin susceptibility and specific heat, respectively, can be written as a Taylor series around $\beta = 0$
\begin{gather}
\chi = \sum_{n=0}^{\infty} c_n \beta^n, \\
C = \sum_{n=0}^{\infty} d_n \beta^n,
\end{gather}
where $\beta$ is the inverse temperature.
We would like to note that in this work we neglect the orbital contribution to the magnetic susceptibility since 
its impact on the critical temperature is expected to be small. 
The first coefficient $c_0 = 0$, which gives the decay  of $\chi$ to zero at an infinite temperature, 
corresponds to Curie law ($c_1$ defines the Curie constant).  
Critical temperature limits the radius of convergence of the series, since at the critical temperature the thermodynamic quantities are not analytic functions.  

To find the critical temperature, the following ratios are commonly used
\begin{gather}
q_n = \left|\frac{c_n}{c_{n-1}}\right|, \\
s_n = \left|\frac{d_n}{d_{n-1}}\right|.
\end{gather}
The $q_n$ and $s_n$ correspond to the d'Alembert test of convergence. 
At a large $n$, both $q_n$ and $s_n$ converge to the critical temperature.  
Another commonly used test is the root test based on $|c_n|^{1/n}$. 
However, each $c_n$ is a homogeneous polynomial of degree $(n-1)$ of $J_1$ and $J_2$. 
So the root estimate is unphysical because it is not strictly proportional to $J$.  
The ratios do not have this problem. 

Instead of applying the root test to the $\chi$, 
we apply the root test to $\frac{\chi}{\beta}$ or $\frac{\partial \chi}{\partial \beta}$: 
\begin{gather}
g_n = | c_n|^{1/(n-1)} \to T_c, n \to \infty \\
h_n = |n c_n|^{1/(n-1)} \to T_c, n \to \infty
\end{gather}
The $g_n$ and $h_n$ are proportional to $J$, which makes them more reliable. 

At the same time, coefficients for the specific heat $d_n$ are homogeneous polynomials of degree $n$ of $J_1$ and $J_2$. 
\begin{gather}
f_n = | d_n|^{1/n} \to T_c, n \to \infty \\
r_n = |n d_n|^{1/n} \to T_c, n \to \infty
\end{gather}
We applied all of the estimators described above, which we report in Section~3 in SI. 
We also tried various extrapolation techniques\cite{Aitken:1927,Lubkin:W-transform:1952}, decomposition of $\log(\chi/\beta)$ and $\log(\chi')$, 
but the results were not satisfactory. 
Therefore, in this work, we limit our consideration to $q_n$, $s_n$, $g_n$, $h_n$, $f_n$, $r_n$.

\textbf{Computational setup.} We followed the computational protocol designed in Ref.~\cite{Pokhilko:BS-GW:solids:2022}
and used unit cells of different types to capture antiferromagnetic (broken-symmetry) solutions of different types. 
The used lattice constants are 4.1705\AA~\cite{Morosin:NiO:exchange_striction:1971}, 
4.4450\AA~\cite{MnO_a}, 4.2630\AA~\cite{Takeuchi:CoO:1979}, 4.285\AA~\cite{Crisan:FeO:2011} 
for NiO, MnO, CoO, and FeO respectively.  
All calculations are performed with \emph{gth-dzvp-molopt-sr} basis set\cite{GTHBasis}, \emph{gth-pbe} pseudopotential\cite{GTHPseudo}, \emph{def2-svp-ri} auxiliary basis\cite{RI_auxbasis} for
the resolution-of-identity decomposition, 
the Monkhorst--Pack k-point grid for the Brillouin-zone sampling\cite{Monkhorst:Pack:k-grid:1976} ($4\times 4\times 4$ for CoO and $5\times 5\times 5$ for FeO), 
Ewald approach\cite{EwaldProbeCharge,CoulombSingular} for the treatment of the finite-size effects, 
and an intermediate representation\cite{Yoshimi:IR:2017} 
with $\Lambda = 10^5$ and 136 functions as a frequency grid.  
We used the one- and two-electron integrals computed with the PySCF code\cite{PYSCF}  
to perform the GW calculations with the local in-house Green's function code \cite{Rusakov16,Iskakov20,Pokhilko:tpdm:2021,Pokhilko:local_correlators:2021,Yeh:GPU:GW:2022,Yeh:X2C:GW:2022}. 
We used the frequency-dependent CDIIS algorithm\cite{Pokhilko:algs:2022} to accelerate the convergence of finite-temperature self-consistent GW iterations.  
For the high-temperature expansion, we wrote a program generating a spin graph with periodic boundary conditions for a cubic lattice and used it to
construct a $20\times 20 \times 20$ Hamiltonian from the Eq.~\ref{eq:Ham_defs}. This Hamilonian was then passed to 
the HTE10 code\cite{Richter:HTE10:code:2014} 
generating expansions up to $10^\text{th}$ order. 
We list the resulting expressions in SI in Section~2.  

\protect\label{sec:main_res}
\begin{table} [tbh!]
  \caption{The effective exchange couplings in K and their ratios employed in this work. 
\protect\label{tbl:Js}}
\makebox[\textwidth]{
\begin{tabular}{c|ccc|c|ccc}
\hline
\hline
            & \multicolumn{3}{c|}{NiO} & & \multicolumn{3}{c}{MnO}  \\ 
            & $J_{1}$  & $J_{2}$ & $J_2/J_1$           & & $J_{1}$ & $J_{2}$ & $J_2/J_1$           \\
\hline
            \multicolumn{8}{c}{Green's function methods}  \\ 
\hline
UHF$^a$         & 9.06   & -57.77  & -6.378 & UHF$^a$ & -2.61 & -2.09 & 0.800 \\
GW$^a$          & 18.95  & -157.96 & -8.335 & GW$^a$  & -7.56 & -6.70 & 0.887 \\
\hline
            \multicolumn{8}{c}{Wave-function methods}  \\ 
\hline
CASSCF$^b$      & 5.80   & -58.02 & -10.00   & CASSCF$^c$ & -31.9    & -- &  \\
CASPT2$^b$      & 13.93  & -193.80 & -13.92 & CASPT2$^c$ & -95.0   & -- &  \\
DDCI2$^b$      & 13.93    & -146.22 & -10.50 &          &     &    &  \\
DDCI3$^b$       & 20.89   & -189.16 & -9.056 &          &     &    &  \\
\hline
\hline
            & \multicolumn{3}{c|}{CoO} & & \multicolumn{3}{c}{FeO}  \\ 
            & $J_{1}$  & $J_{2}$ & $J_2/J_1$           & & $J_{1}$ & $J_{2}$ & $J_2/J_1$           \\
\hline
            \multicolumn{8}{c}{Green's function methods}  \\ 
\hline
UHF$^d$         & 3.13   & -11.42  & -3.647     & UHF$^d$ & 3.64  & -5.62  & -1.545  \\
GW$^d$          & 3.30   & -33.41  & -10.12     & GW$^d$  & 3.83  & -13.56 & -3.540  \\
\hline
\end{tabular}}

$^a$: Reference\cite{Pokhilko:BS-GW:solids:2022} 

$^b$: Reference\cite{Martin:NiO:exchange:2002} 

$^c$: Reference\cite{deGraaf:MRPT:exchange:solids:SMM:2001} 

$^d$: This work. 
\end{table}

For the transition-metal oxides studied, Table~\ref{tbl:Js} shows the effective exchange couplings $J$  evaluated with rigorous ab-initio methods without any adjustable parameters.  
The broken-symmetry UHF and GW estimates for FeO and CoO evaluated in this work yield 
ferromagnetic $J_1$ and antiferromagnetic $J_2$ constants. 
For both NiO and MnO, similarly to FeO, GW increases the magnitude of $J_2$ in comparison to UHF.
This can be explained by enhancement of the superexchange due to a partial stabilization of charge-transfer contributions because of the effective screened interaction $W$. 
\begin{figure}[!h]
  \includegraphics[width=8cm]{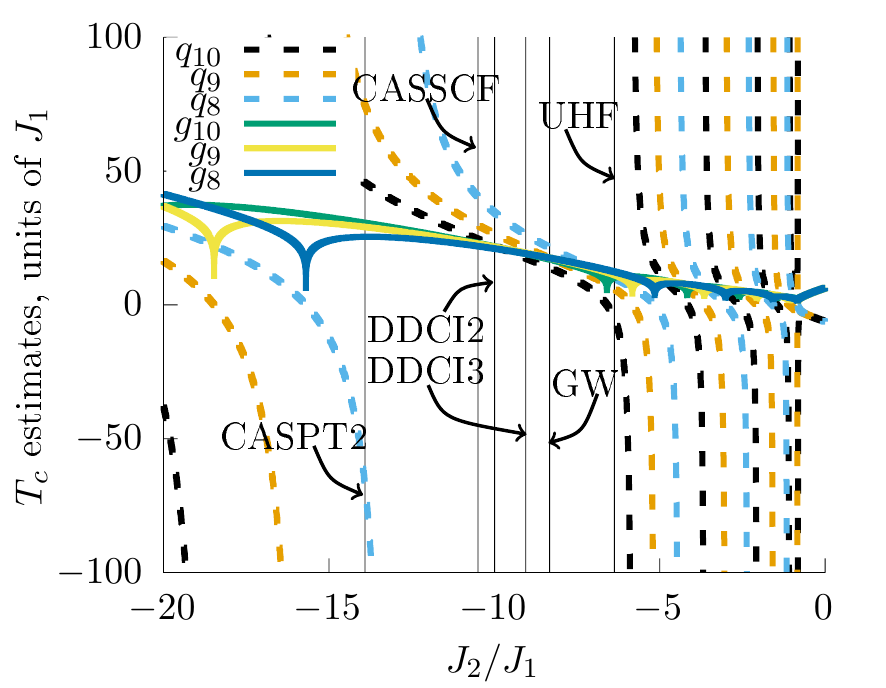}
\centering
\caption{Behavior of various estimates with respect to the $J_2/J_1$ ratios for NiO. 
         The shown $q_n$ ratios are signed to show the pole structure. 
         Specific values of the $J_2/J_1$ ratios from quantum chemical calculations are marked with vertical lines. 
         Neel temperature estimates can be seen as intersection points between the vertical lines and 
         the ratio or root estimates. 
         \protect\label{fig:NiO_graph}}
\end{figure}

We validate the high-temperature expansion by applying it to cubic lattices as described in 
Section~1 in SI.  There, we also provide a comparison of limiting cases with the previously published results from Refs\cite{Rushbrooke:Wood:HTE:1955,Domb:Sykes:HTE:Curie:1957,Wessel:QMC:crit:cubic:2004,Wood:HTE:NNN:Curie:1967}. 

The Neel temperatures extracted from different convergence radius estimates can vary vastly. 
This behaviour is illustrated by Figure~\ref{fig:NiO_graph}, 
which shows the dependence of $q_n$ and $g_n$ on the ratio of $J_2$ and $J_1$ constants for $S=1$ case, 
corresponding to NiO. 
The $q_n$ ratios change smoothly with $J_2/J_1$ for positive $J_2/J_1$, but show multiple poles 
when $J_2/J_1$ is negative.  
The $J_2/J_1$ ratios computed with quantum chemical calculations lie in close proximity to the $q_n$ poles 
resulting in a high sensitivity of $q_n$ on the $J_2/J_1$. 
The estimates based on the roots (Table~S3 in SI) are much more stable and show a clear convergence pattern.  

The behavior of the convergence radius estimates for CoO and FeO ($S=3/2$ and $S=2$) is similar to NiO. 
In SI, Tables~S4 and S5 show the root estimates 
linking the Neel temperature and $J_2$ for a given $J_2/J_1$ values.  
An increase in $S$ leads to higher $T_N / J_2$ values, which is expected from the form of the polynomials 
as explained in Refs\cite{Richter:HTE8:2011,Richter:HTE10:code:2014}.  
\begin{figure}[!h]
  \includegraphics[width=8cm]{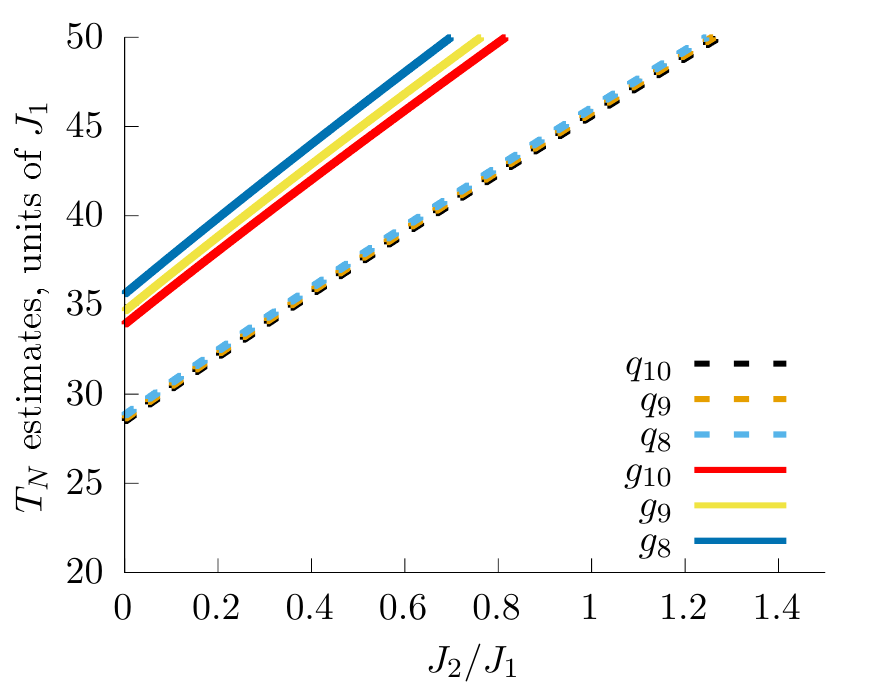}
\centering
\caption{Behavior of various estimates with respect to the $J_2/J_1$ ratios for MnO. 
         \protect\label{fig:MnO_graph}
}
\end{figure}

In SI, Tables~S6 and S7 show the estimates for MnO. 
Since all the poles of $q_n$ are located in the negative $J_2/J_1$ region, 
the positive $J_2/J_1$ provided by UHF and GW are far enough from the poles to stabilize the estimates based on $q_n$ 
(Figure~\ref{fig:MnO_graph}). 
Indeed, all $q_n$ converge faster than the root estimates, leading to a good Domb--Sykes extrapolation (Table~S1 in SI).   
The multiplication constants connecting the $T_N$ and $J_2$ for MnO are much larger than for other compounds, 
which cannot be rationalized by the $S$ values alone.  
Since the $J_2/J_1 \approx 1$, both NN and NNN are coupled with comparable constants, 
resulting in an effective coordination number equal to $18$. 
In the mean-field treatment, 
the critical temperature is proportional to the coordination number. 
Early publications on high-temperature expansions\cite{Rushbrooke:Wood:HTE:1955,Domb:Sykes:HTE:Curie:1957} preserved this trend and showed that the larger the coordination number is, the larger the multiplier is, which explains our findings. 
\begin{table} [tbh!]
  \caption{Estimates of Neel temperature (K) based on $g_{10}$ and $h_{10}$, $f_{10}$ and $r_{10}$.   
\protect\label{tbl:Neel_T}}
\begin{tabular}{c|c|c|c|c|}
\hline
\hline
      & NiO & CoO & FeO & MnO\\
Exp   & 530$^a$, 516$^b$, 524.5$^c$, 523$^d$, 520$^g$    & 288$^e$, 293$^g$, 289.7$^g$  & 198$^g$, 183$^g$    & 118$^f$, 122$^g$, 116$^g$  \\
\hline
UHF   & 94--121; 129--162   & 48--62; 53--66     & 36--46; 47--59   & 107 ($q_\infty$); 100 ($s_\infty$)    \\
GW    & 324--419; 349--439  & 153--198; 144--181 & 70--90; 103--130 & 321 ($q_\infty$); 299 ($s_\infty$)  \\
\hline
DDCI2 & 320--414; 320--403  &     &     &    \\
DDCI3 & 401--518; 416--524  &     &     &    \\
CASSCF & 126--163; 127--160 &    &     &    \\
CASPT2 & 426--551; 419--528 &    &     &    \\
\hline
\end{tabular}

$^a$: Reference\cite{Lindgard:NiO:2009}

$^b$: Reference\cite{Vernon:NiO:1970}

$^c$: Reference\cite{Seehra:NiO:1984}

$^d$: Reference\cite{Balagurov:NiO:MnO:2016}

$^e$: Reference\cite{Dhalenne:NiO:CoO:critical:2008}

$^f$: Reference\cite{Goncharenko:MnO:FeO:neutron_dif:2005}

$^g$: Reference\cite{Kubo:antiferromagnetism:1955}
\end{table}

Table~\ref{tbl:Neel_T} shows Neel temperatures extracted from the experimental measurements 
and evaluated from ab-initio calculations. 
For NiO, DDCI3 provides the most reliable estimate of $J_1$, $J_2$, and consequently the Neel temperature that is closest to experimental data. 
GW $J_2/J_1$ ratio is close to the DDCI3, which indicates that GW provides a balanced estimate of 
the effective exchange constants. 
GW improves the treatment of electronic correlation from UHF and provides estimates of the exchange constants 
close to DDCI2, which explains that a substantial part of dynamic electronic correlation in these compounds comes from 
screening.  
Underestimation of the absolute values of the GW exchange constants result in an underestimation of the Neel temperature. 
All the GW root estimates reproduce the trend in the experimental estimates of Neel temperatures for NiO, CoO, and FeO. 
An agreement of FeO and CoO with this trend indicates that the spin--orbit interaction does not 
distort the structure of the magnetic Hamiltonian significantly to change the effective description from $S=2$ and $S=3/2$ 
to lower values, 
which is consistent with Kanamori's prediction\cite{Kanamori:FeO:CoO:I:1957}.  
This observation is contrary to Refs.\cite{Fink:CoO:2002,Stock:CoO:2020} where only the lower-lying manifold of spin--orbit--adiabatic states 
were considered for the effective magnetic Hamiltonian.  

However, the MnO is different. 
Surprisingly, the UHF Neel temperature estimates are close to the experimental estimates for MnO, 
which is likely due to a fortuitous error cancellation. 
The GW increases the strengths of magnetic interactions and overestimates the Neel temperature in MnO. 
A plausible reason behind this disagreement is the existence of a structural phase transition from the low-temperature distorted structure 
to the high-temperature rock-salt structure. 
According to neutron diffraction experiments\cite{Goncharenko:MnO:FeO:neutron_dif:2005,Balagurov:NiO:MnO:2016}, 
in NiO and FeO this structural transition happens at a much lower temperature than the magnetic phase transition. 
In MnO, the structural and magnetic phase transitions either happen together or separated only by $\sim 1$~K. 
In either case, one can expect substantial spin-vibronic effects\cite{Cottam:cubic:oxides:2021}, 
which are not taken into account in our calculations.

In this work, we applied broken-symmetry fully self-consistent GW approach and evaluated the effective exchange couplings 
in nickel, manganese, cobalt, and iron oxides. 
In both FeO and CoO, the $J_1$ constant is ferromagnetic and the $J_2$ constant is antiferromagnetic similar to NiO.  
The GW estimate of $|J_2|$ is several times larger than the UHF estimate. 
This observation is consistent with other solid and molecular compounds with significant superexchange, 
which can be explained by the inclusion of screened interactions in GW stabilising ionic contributions\cite{Pokhilko:local_correlators:2021,Pokhilko:BS-GW:solids:2022}.  
The UHF and GW estimates of $J_1$ are similar, a possible explanation of which 
is in the dominance of the direct mechanism of exchange.  

We found that when the signs of $J_1$ and $J_2$ are different, the $q_n$ ratios contain multiple poles close to the computed $J_1/J_2$ values, making critical temperature estimates based on $q_n$ unreliable in this regime. 
The estimates based on roots ($g_n, h_n, f_n, r_n$) are much more stable and converge faster than $q_n$. 
The Neel temperature estimates derived from spin susceptibility and specific heat are 
in excellent agreement with each other.  
 For NiO, the critical temperature evaluated from GW is in agreement with the estimates 
from the wave-function calculations.  
The estimated critical temperatures from GW for NiO, CoO, and FeO reproduce the trend in the experimental estimates of critical temperature for this series of compounds. 

If the signs of $J_1$ and $J_2$ are the same, the $q_n$ are separated from the poles and become stable.  
This regime is observed for MnO, for which $q_n$ converge faster than the root estimates.  
The Neel temperature of MnO is very sensitive to its $J$ values, which can be explained by large spin ($S=5/2$) 
and large effective coordination number. 
For MnO, the GW overestimates the critical temperature. 
While the precise cause of this overestimation is not known, 
plausible causes, such as spin-vibronic effects and deviation from Heisenberg Hamiltonian,  
will be investigated in our future work.

\clearpage

\section*{Acknowledgments}
D.Z. and P.P. were supported by the Simons Foundation via the
Simons Collaboration on the Many-Electron problem.

\section*{Supplementary Material}
Validation, expressions for HTE coefficients, Neel temperature estimates.

\renewcommand{\baselinestretch}{1.5}

\end{document}